\documentclass[12pt]{revtex4}

\usepackage{amssymb, amsmath, amsfonts, latexsym, verbatim, mathrsfs}

\usepackage{graphicx}

\newtheorem{theorem}{Theorem}

\begin{document}

%\markboth{GianCarlo Ghirardi, Raffaele Romano}
%{Classical, quantum and superquantum correlations}
%
%%%%%%%%%%%%%%%%%%%%%% Publisher's Area please ignore %%%%%%%%%%%%%%%
%%
%\catchline{}{}{}{}{}
%%
%%%%%%%%%%%%%%%%%%%%%%%%%%%%%%%%%%%%%%%%%%%%%%%%%%%%%%%%%%%%%%%%%%%%%

\title{CLASSICAL, QUANTUM AND SUPERQUANTUM CORRELATIONS }

\author{GianCarlo Ghirardi}
\email{ghirardi@ictp.it}
\affiliation{Department of Physics, University of Trieste, the Abdus Salam ICTP, Trieste  \\
Strada Costiera 11, I-34151 Trieste, Italy}

\author{Raffaele Romano}
\email{rromano@ts.infn.it}
\affiliation{Department of Physics, University of Trieste, Fondazione Parisi, Rome, Italy}

\begin{abstract}
A deeper understanding of the origin of quantum correlations is expected to shred light on the physical principles underlying quantum mechanics. In this work, we reconsider the possibility of devising ``crypto-nonlocal theories", using a terminology firstly introduced by Leggett. We generalize and simplify the investigations on this subject which can be found in the literature. At their deeper level, such theories allow nonlocal correlations which can overcome the quantum limit.
\end{abstract}

%\begin{history}
%\received{Day Month Year}
%\revised{Day Month Year}
%%\accepted{(Day Month Year)}
%%\comby{(xxxxxxxxxx)}
%\end{history}


\maketitle

%%%%%%%%%%%%%%%%%%%%%%%%%%%%%%%%%%%%%%%%%%%%%%%%%%%%%%%%%%%%%%%%%%%%%%%%

\section{Introduction}

The quantum correlations between the far-away constituents of a composite system in an entangled state are probably the most characteristic, interesting and puzzling aspect of modern physics. In the infancy of quantum theory they have been considered as a weird feature of the formalism, as clearly stressed by Schr\"{o}dinger himself\cite{Erwin}:
\begin{quote}
{\it It is rather discomforting that the theory should allow a system to be steered or piloted into one or the other type of state at the experimenter's mercy in spite of his having no access to it}
\end{quote}
This sentence makes clear that the really puzzling aspects of the theory derive from its calling into play, in a way or another, non local features. Precisely in the same year in which the above sentence was written, the celebrated EPR paper was published\cite{EPR}. It represents an extremely important step in giving the due relevance to the question of locality. Its content can be summarized in the following oversimplified way: any theory {\it which is local} and reproduces the quantum correlations determined by the assignment of the state vector must be recognized as incomplete, i.e. it requires further formal elements besides the wavefunction to completely characterize the state of an individual physical system.

In 1935, neither Einstein, nor scientists like Bohr, Heisenberg and Born, were inclined to  give up locality and this fact gave origin to two different positions: those who shared the EPR view  entertained the idea of completing the formalism either by the addition to (or the replacement of) the state vector of (with) further variables, denoted as hidden variables. On the other hand the enourmous influence of the ``orthodox" position has led the large majority of the scientific community to firmly believe that  quantum theory,  the ultimate theory of nature, gives a complete description of natural processes. Accordingly, the puzzling aspects of nonlocality have been exorcized either by rather obscure arguments as those of Bohr in his reply to the EPR paper, or by facing directly the completeness problem trying to prove that a completion of the theory was impossible. Actually it was von Neumann\cite{von} who gave an enormous boost to this idea by deriving his impossibility proof of a completion of the theory. His argument was formally correct but, as well known now, it made use of a logically non necessary assumption (which became clear after the lucid analysis by Bell\cite{Bell1}) and consequently it had no conceptual relevance. One might state that the  prestige of von Neumann has discouraged any attempt of working out a hidden variable theory for about twenty years.

In 1952 D. Bohm\cite{Bohm} exhibited his celebrated model, a deterministic completion of quantum mechanics, predictively equivalent yet radically different from it (for instance, in this theory all particles possess precise positions at all times). Obviously, in accordance with the EPR conclusions, Bohmian mechanics exhibits precise nonlocal aspects. At this point J. Bell enters the game and proves, by deriving his celebrated inequality\cite{Bell2}, that any complete theory whatsoever which reproduces the (now well tested) quantum correlations must violate a quite natural locality request we will make formally precise in what follows.

It seems also useful to mention that assuming locality is a quite natural request within the classical conception of natural processes since, due to the deterministic character of classical theories, its violation would give rise to a direct conflict with relativistic requirements.

In this paper, we will discuss the problem of the correlations between separate parts of a composite system by taking the two views which are possible concerning their (local or nonlocal) nature. As it is usual, we will introduce an appropriate combination (which we will denote here as $F$) of the correlation functions referring to different measurement processes and we will analyze it first of all by assuming locality. In this case it is possible to derive the upper bound of 2 (the bound which embodies the celebrated CHSH inequality\cite{CHSH}) for $F$ . Subsequently, we release the locality request and we  discuss the bound to $F$ which can be derived in general (i.e. for arbitrary nonlocal correlations subjected only to the request that they do not allow superluminal signaling) and within the specific framework of quantum theory (i.e. by taking into account the values of the quantum correlations). Accordingly, we will analyze the so-called {\it Popescu-Rohlrich box}\cite{Popescu}, a very smart device that produces the strongest possible nonlocal correlations, i.e. it allows $F$ to reach its maximum possible value of 4. We will also spend some time to analyze the nonlocal structure of the PR box by taking the perspective which has been introduced by Jarrett\cite{Jarrett} and Shimony\cite{Shimony} within the quantum framework. Having done this, we  pass to discuss the same problem in a strictly quantum context, and we recall that the bound in such a case is the one given by the Tsirelson inequality\cite{Tsirelson}, and the maximal violation which can occur turns out to be $2\sqrt{2}$.

The fact that the quantum bound does not saturate the general nonlocal bound is rather surprising and, in our opinion, it deserves a deeper analysis. We are thus  led to investigate the interesting class of {\it crypto-nonlocal hidden variable theories} recently introduced by Leggett\cite{Leggett}. Such theories contemplate the possibility of two levels of hidden variables in such a way that the local average over the deeper level washes out nonlocality. The conclusion we will reach is quite interesting and might help in throwing some light on the deep question of nonlocality. Actually we will exhibit a very simple crypto-nonlocal model which, while exhibiting superquantum correlations (i.e., just as the PR box it violates Bell's bound) when the average is taken on the lower lying variables, reproduces quantum mechanics, and consequently it respects Bell's bound, when the averages involve also the upper level variables\cite{Ghirardi1}.

\section{Quantum correlations and nonlocality}

We consider a bipartite system, with two-dimensional far-away\footnote {By the expression far away we intend to synthetically summarize the fact that we assume that the whole series of events, choice of the set-ups of the measuring devices, actual performing and completion of the measurements and reading of the outcomes take place in space-like separated regions.}   constituents. Let $A({\bf a})$, $A({\bf a}^{\prime})$, $B({\bf b})$ and $B({\bf b}^{\prime})$ be local observables which can assume the values $-1$ and $+1$, with ${\bf a}$ and ${\bf a}^{\prime}$ representing the settings (e.g., spin directions) characterizing the measurements performed by Alice, and ${\bf b}$ and ${\bf b}^{\prime}$ having the same meaning for the Bob's side. We denote as $P(A, B \vert {\bf a}, {\bf b})$ the joint probability of obtaining $A({\bf a}) = A$ and $B({\bf b}) = B$ when both $A({\bf a})$ and $B({\bf b})$ are measured, a quantity which obviously depends on the physical model which describes the bipartite system.

In the following, we will refer to the following notion of locality: a physical model producing the aforementioned joint probabilities is said to be local if and only if these probability factorize,
\begin{equation}\label{loc}
    P(A, B \vert {\bf a}, {\bf b}) = P(A \vert {\bf a}) P(B \vert {\bf b}),
\end{equation}
for every outcomes $A, B$, and every possible choice of settings ${\bf a}, {\bf b}$. A theory which satisfies (\ref{loc}) is called a local theory. It has been proven that this notion amounts to the logical conjunction of two other requests which have been named {\it Locality} and {\it Completeness}, respectively, by Jarret\cite{Jarrett} and {\it Parameter Independence} (PI) and {\it Outcome Independence} (OI), by Shimony\cite{Shimony}. Outcome Independence is expressed by:
\begin{equation}\label{outin}
P(A|{\bf a},{\bf b};B) = P(A|{\bf a},{\bf b}), \qquad P(B|{\bf a},{\bf b};A) = P(B|{\bf a},{\bf b}),
\end{equation}
that is, the outcome of the measurement on one side does not affect the probability of the outcome on the other side. Parameter independence requires that
\begin{equation}\label{parin}
P(A|{\bf a},{\bf b}) = P(A|{\bf a}), \qquad P(B|{\bf a},{\bf b}) = P(B|{\bf b}),
\end{equation}
that is, the measurement setting on one side does not affect the probability of the outcome of the measurement  on the other side. By using the Bayes rule,
\begin{equation}\label{bayes}
P(A, B|{\bf a},{\bf b}) = P(A|{\bf a},{\bf b};B)P(B|{\bf a},{\bf b}),
\end{equation}
and considering (\ref{outin}) and (\ref{parin}), one finds the locality condition (\ref{loc}). On the other hand it is trivial to go the other way around showing that this last condition implies both OI and PI. Therefore, locality is equivalent to the joint requirements of OI and PI.

In the study of the locality issue, it is convenient to make use of some marginals of the joint probabilities. The correlation $E({\bf a},{\bf b})$ of the outcomes obtained by the two parties is given by
\begin{equation}\label{cor}
E({\bf a},{\bf b}) = P(A = B \vert {\bf a},{\bf b}) - P(A \ne B \vert {\bf a},{\bf b}),
\end{equation}
and it represents the average value of the product of the outcomes. In full generality, Clauser, Horne, Shimony and Holt\cite{CHSH}, by deepening the analysis of Bell\cite{Bell2}, have shown that an appropriate combination of such correlations (with ${\bf a},{\bf b},{\bf a}^{\prime}$ and ${\bf b}^{\prime}$  arbitrarily chosen), i.e.;
\begin{equation}
F \equiv E({\bf a},{\bf b}) + E({\bf a},{\bf b}^{\prime}) + E({\bf a}^{\prime},{\bf b}) - E({\bf a}^{\prime},{\bf b}^{\prime}),
\end{equation}
satisfies, for all local theories, the inequality (named CHSH inequality after them),
\begin{equation}\label{CHSH}
\vert F \vert \leqslant 2.
\end{equation}
On the other hand, in the quantum case, when consideration is given to two far away spin $1/2$ particles in the singlet state
\begin{equation}\label{singlet}
    \vert \phi_- \rangle = \frac{1}{\sqrt{2}} (\vert 01 \rangle - \vert 10 \rangle),
\end{equation}
and measurements of the spin components along appropriate directions are performed, the correlations are given by
\begin{equation}\label{qcor}
    E_{\phi_-}({\bf a},{\bf b}) \equiv \langle \phi_- \vert \hat{A}({\bf a}) \otimes \hat{B}({\bf b}) \vert \phi_- \rangle,
\end{equation}
where $\hat{A}({\bf a})$, $\hat{B}({\bf b})$ are the Hermitian operators which quantum mechanics associates to $A({\bf a})$ and $B({\bf b})$ respectively.
It is possible to prove that these correlations violate the above inequality for appropriate choices of ${\bf a},{\bf b},{\bf a}^{\prime}$ and ${\bf b}^{\prime}$. Any hidden variable theory predictively equivalent to quantum mechanics has to do the same, therefore it has to be nonlocal.
Actually, for the correlations exhibited by a generic quantum state $\psi$, Tsirelson has derived the analogous of the CHSH inequality\cite{Tsirelson}, expressed by
\begin{equation}\label{tsirel}
\vert F_{\psi} \vert \leqslant 2 \sqrt{2}, \qquad F_{\psi} \equiv E_{\psi}({\bf a},{\bf b}) + E_{\psi}({\bf a},{\bf b}^{\prime}) + E_{\psi}({\bf a}^{\prime},{\bf b}) - E_{\psi}({\bf a}^{\prime},{\bf b}^{\prime}),
\end{equation}
where $E_{\psi}({\bf a},{\bf b})$ has the form (\ref{qcor}), with $\phi_-$ replaced by $\psi$. The right hand side upper bound can actually be reached for appropriate choices of the observables appearing in it, if the system is in a maximally entangled state (e.g., the aforementioned singlet state). This rises an interesting question: is it possible to exhibit nonlocal models which violate the Tsirelson bound, exactly as the CHSH bound is violated by quantum mechanics?
Can these models have any impact on our understanding of quantum non-locality, and, in particular, to clarify what are the physical principles underlying it?

\section{Superquantum correlations}

As we have seen, the correlations between the outcomes of measurements performed on far away quantum systems in entangled states exhibit an irreducible and unavoidable nonlocal nature. Such nonlocal correlations do not conflict with relativistic causality, since they cannot be used for superluminal communication.
Nevertheless, a general nonlocal theory could be inconsistent with this principle, depending on its specific structure. One might further expect that the combination of correlations in (\ref{cor}) could reach the value $4$, which is attained when the first three terms take the value $+1$ and the last one the value $-1$.

Popescu and Rohrlich\cite{Popescu} have analyzed in detail what constraints have to be satisfied to avoid any conflict with relativistic causality. Their analysis starts by raising the question of whether this request could be responsible of the specific form that nonlocality takes in quantum mechanics. They have answered in the negative, by introducing the so-called {\it PR box}, a device with classical bits as inputs and outputs, which gives rise to correlations which violate the Tsirelson bound (\ref{tsirel}) to the maximal extent, but cannot be used for faster-than-light signalling. In general, correlations which violate the Tsirelson bound are called {\it superquantum correlations}.

We analyze in some detail the PR box. Its working is summarized by the following relation between the inputs $x$, $y$ and the outputs $a$, $b$, each of which is assumed to take only the values $0$, $1$:
\begin{equation}\label{PR box}
a + b = xy \quad {\rm mod} \; 2.
\end{equation}
Following \cite{Cerf}, it is convenient to redefine the outcomes as $a'$ and $b'$ such that their possible values are $-1$ and $+1$. They are related to $a$ and $b$ by
\begin{equation}\label{newout}
a' = 1 - 2a, \qquad b' = 1 - 2b.
\end{equation}
We now consider the correlations $E_{PR}(x,y)$ characterizing the PR box,  defined as in (\ref{cor}), which are easily calculated. When $x = y = 0$ one must have $a = b$, and then $a' = b'$, and one gets $E_{PR}(0,0) = 1$. The same conclusion holds  for $x = 0$, $y = 1$ and for $x = 1$, $y = 0$. Finally, when $x = y = 1$ the outcomes $a$ and $b$ are different so that only $P(+1,-1 \vert 1,1)$ and $P(-1,+1 \vert 1,1)$ contribute to $E_{PR}(1,1)$. But these probabilities sum to $1$, and then $E_{PR}(1,1) = -1$. The general combination expressing the violation of locality reads:
\begin{equation}\label{sat4}
F_{PR} \equiv E_{PR}(0,0) + E_{PR}(1,0) + E_{PR}(0,1) - E_{PR}(1,1) = 4,
\end{equation}
and, as stated before, it saturates the upper bound of (\ref{cor}). If the inputs of the PR box are random, the same is true for its outputs. Therefore, the device cannot be used to signal, and it respects relativistic causality.

We now investigate in more detail the features of nonlocality exhibited by the PR box. We consider two possible scenarios: (i) we describe the device only by  (\ref{PR box}); (ii) we consider an hidden variable model which reproduces the inputs-outputs relations. In both cases, we study the properties of Parameter and Outcome Independence.

(i) Suppose that  Alice's input $x$ is known. Then, if $x = 0$, it follows that $x y = 0$, that is $(a = 0, b = 0) \vee (a = 1,b = 1)$. So, the outcome $a$ that Alice obtains once she knows her setting $x = 0$ depends on the outcome obtained by Bob, and OI is violated. On the contrary, if $x = 1$ two cases are possible: either $y = 0$ and we are back to the previous situation, or $y = 1$, producing $(a = 1, b = 0) \vee (a = 0,b = 1)$. In this case, the joint knowledge of $b$ and of $y$ is necessary to determine $a$. Therefore, given the  input by Alice, the corresponding output depends, in general, from both $y$ and $b$: the theory violates both PI as well as OI. Completely analogous considerations hold for the setting and the outcome of Bob.

It is possible to face the problem by looking at the product $ab$ of the outcomes. If both settings of Alice and Bob are given, there are two possibilities: if $xy = 0$ we know that for sure $a = b$ but we do not know their specific value. Alternatively, if $xy = 1$ we know that $a$ and $b$ must be different, but, once more, we do not know the actual value of them. Specification of both settings does not determine the outcomes. Some further knowledge is necessary.

(ii) Suppose now we consider a hidden variable model characterized by the variable $\lambda$, which also can take the values $0$, $1$ and, to be completely general, let us assume that the probabilities of its two values are given by $P(\lambda = 0)$ and $P(\lambda = 1)$. Obviously, $P(\lambda = 0) + P(\lambda = 1) = 1$. The model is defined by the assumption that, for any given settings $x$ and/or $y$ for Alice and Bob respectively, the assignment of $\lambda$ determines the  outcome(s) according to the following rules:
\begin{equation}\label{prhidden}
a = (x + \lambda) \quad  {\rm mod} \; 2, \qquad
b = (x + \lambda - x y) \quad {\rm mod} \; 2.
\end{equation}
Note that the model is manifestly nonlocal since the value of the outcome $b$ besides depending on the value of the hidden variable $\lambda$ and of the  input $y$ depends also on the input $x$. In accordance with (\ref{prhidden}) the outputs are fixed as in Table \ref{tab1}.
%\begin{table}[ph]   %Table~1
%\tbl{This is the caption for the table. If the caption is less than
%one line then it is centered. Long captions are justified to the table
%width by using the command $\backslash$tbl.}
%{\begin{tabular}{@{}lll@{}} \Hline
%\\[-1.8ex]
%Schedule & Capacity &
%Level \\[0.8ex]
%\hline \\[-1.8ex]
%Business plan & Financial planning &
%Planning\\ Production planning & Resource requirement plan (RRP) &{}\\
%Final assembly schedule & Capacity control &{} \\
%Master production schedule (MPS) & Rough cut capacity plan (RCCP) &{}\\
%Material requirement plan & Capacity requirement plan (CRP) &{}\\
%Stock picking schedule & Inventory control & {} \\
%Order priorities & Factory order control & Execution \\
%Scheduling & Machine (work-centre) control & {}\\
%Operation sequencing & Tool control$^{\rm a}$ & {} \\[0.8ex]
%\Hline \\[-1.8ex]
%\multicolumn{3}{@{}l}{$^{\rm a}$ Table footnote.}\\
%\end{tabular}}
%\label{tab1}
%\end{table}
\begin{table}[htdp]
\caption{Outcomes as functions of the inputs and of $\lambda$.}
{\begin{tabular}{c|c|c||cc}
\quad x \quad & \quad y \quad & \quad $\lambda$ \quad & \quad a \quad & \quad b \quad \\ [0.5ex]  \hline
0 & 0 & 0 & 0 & 0 \\
0 & 0 & 1 & 1 & 1 \\
1 & 0 & 0 & 1 & 1 \\
1 & 0 & 1 & 0 & 0 \\
0 & 1 & 0 & 0 & 0 \\
0 & 1 & 1 & 1 & 1 \\
1 & 1 & 0 & 1 & 0 \\
1 & 1 & 1 & 0 & 1
\end{tabular}}
\label{tab1}
\end{table}%
Looking at the table one immediately checks that the basic relation characterizing the inputs and outputs of the PR box, Eq. (\ref{PR box}), is satisfied.
The two outputs depend on the inputs of the other party (nonlocality) but  not  on the associated output. We can then claim that the model exhibits PI (as any deterministic nonlocal hidden variable model must do) but not OI.

We conclude that the specific features of nonlocality of the PR box are highly dependent on the model which is considered for the description of the device. This is analogous to the analysis of nonlocality of the singlet state (\ref{singlet}), which can be explained as violation of OI, as in the standard quantum mechanical description, or can involve PI, if a deterministic hidden variable model is used to model the system.

A specific model, which relies on the use of a PR box and additional hidden variables, is outlined here for further reference. It makes evident that it is possible to simulate the quantum correlations by using the PR box. The hidden variables are two three-dimensional real vectors, $\lambda_1$ and $\lambda_2$, uniformly and independently distributed on the unit sphere. If Alice and Bob measure the spin projections along directions ${\bf a}$ and ${\bf b}$ respectively, and their outcomes can be 0 or 1, they are defined by
\begin{equation}\label{cerf}
  A = a + \frac{1}{2} \big({\rm sgn} ({\bf a} \cdot \lambda_1) + 1\big), \qquad
  B = b + \frac{1}{2} \big({\rm sgn} ({\bf b} \cdot \lambda_+) - 1\big),
\end{equation}
where the Popescu-Rohrlich device has outputs $a$ and $b$ and inputs
\begin{equation}\label{cerfin}
    x = \frac{1}{2}\big({\rm sgn} ({\bf a} \cdot \lambda_1) + {\rm sgn} ({\bf a} \cdot \lambda_2)\big) + 1, \qquad y = \frac{1}{2}\big({\rm sgn} ({\bf b} \cdot \lambda_+) + {\rm sgn} ({\bf b} \cdot \lambda_-)\big) + 1,
\end{equation}
all the sums are modulus 2, and $\lambda_{\pm} = \lambda_1 \pm \lambda_2$ \footnote{Notice that the expressions which define this model are different from those of \cite{Cerf}, where the function ${\rm sgn}$ has range $0, 1$.}. The proof that this model fully reproduces the correlations of the singlet state can be found elsewhere\cite{Cerf,Toner}.

\section{Crypto-nonlocal models}

Some years ago, A. Leggett\cite{Leggett} has paid a specific attention to a quite relevant problem. Having perfectly clear that Bell's inequality implies an irreducible and unavoidable nonlocal nature of physical processes involving far away systems in entangled states, he has been led to discuss whether it might be possible to save some form of "weak locality" without entering in conflict with quantum predictions.

His argument can be synthetically summarized by making reference to the rotationally invariant maximally entangled state of two photons:
\begin{equation}\label{mepol}
\vert \psi_+ \rangle = \frac{1}{\sqrt{2}}[\vert V_{1} V_{2} \rangle + \vert H_{1} H_{2} \rangle].
\end{equation}
Here, $\vert V_{i} \rangle$ and $\vert H_{i} \rangle$ are states of vertical and horizontal plane polarization, respectively. Then one chooses to account for the outcomes of polarization measurements on the system in terms of a nonlocal hidden variable model, characterized by variables of two kinds, $\mu$, with a probability distribution function $\rho_{{\bf u},{\bf v}}(\mu)$, and $({\bf u},{\bf v})$, where ${\bf u}$ and ${\bf v}$ are real three dimensional vectors of length one, distributed according to $F({\bf u},{\bf v})$. The assignment of the hidden variables is assumed to determine uniquely and nonlocally the outcomes $+1,-1$ of linear polarization measurements along any two chosen directions ${\bf a}$ and ${\bf b}$. The outcomes of these measurements are denoted by $A({\bf a,b},\mu)$ and $B({\bf a,b},\mu)$.

Obviously, the quantum mean values for single and joint measurements must coincide with those obtained by averaging over the whole set of hidden variables the precise values assigned to the observables within the hypothetical model. Recalling the definition (\ref{qcor}), and extending it to local averages,
\begin{eqnarray}\label{qcor2}
    && E_{\psi_+}({\bf a},{\bf b}) \equiv \langle \psi_+ \vert \hat{A}({\bf a}) \otimes \hat{B}({\bf b}) \vert \psi_+ \rangle, \\
    && E_{\psi_+}({\bf a}) \equiv \langle \psi_+ \vert \hat{A}({\bf a}) \vert \psi_+ \rangle, \quad E_{\psi_+}({\bf b}) \equiv \langle \psi_+ \vert \hat{B}({\bf b}) \vert \psi_+ \rangle, \nonumber
\end{eqnarray}
we impose
\begin{eqnarray}\label{leggcons}
E_{\psi_+}({\bf a}) &=& \int d {\bf u} \, d {\bf v} \, F({\bf u},{\bf v}) \int d \mu \rho_{{\bf u},{\bf v}}(\mu) A({\bf a,b},\mu), \nonumber \\
E_{\psi_+}({\bf b}) &=& \int d {\bf u} \, d {\bf v} \, F({\bf u},{\bf v}) \int d \mu \rho_{{\bf u},{\bf v}}(\mu) B({\bf a,b},\mu), \\
E_{\psi_+}({\bf a},{\bf b}) &=& \int d {\bf u} \, d {\bf v} \, F({\bf u},{\bf v}) \int d \mu \rho_{{\bf u},{\bf v}}(\mu) A({\bf a,b},\mu) B({\bf a,b},\mu).
\end{eqnarray}

When coming to explicitly impose further requirements to the model, Leggett assumed that the violation of locality did respect Outcome Independence. Actually, this request is necessary because deterministic hidden variable theories can violate Bell's locality condition only by violating PI. We pass now to describe the fundamental condition imposed to the model, denoted by Leggett as {\it crypto-nonlocality}. This amounts to the request that averaging the single particles values for the two observables over the ``deep level" hidden variables $\mu$, nonlocality is washed out,
\begin{eqnarray}\label{cnlhv0}
 \int d \mu \rho_{{\bf u},{\bf v}}(\mu) A({\bf a,b},\mu) = f({\bf a},{\bf u}) \nonumber \\
 \int d \mu \rho_{{\bf u},{\bf v}}(\mu) B({\bf a,b},\mu) = g({\bf b},{\bf v}).
\end{eqnarray}
The reason for this choice derives from a desire to account for the process in terms of an appropriate random emission by the source of pairs of photons with definite plane polarization along ${\bf u}$ and ${\bf v}$. Actually, it is just with this idea in mind that Leggett has chosen a very specific and physically meaningful expression for the functions $f({\bf a},{\bf u})$ and $g({\bf b},{\bf v})$, given by the Malus law,
\begin{equation}
    f({\bf a},{\bf u}) = 2({\bf u} \cdot {\bf a})^{2} - 1, \qquad  g({\bf b},{\bf v}) = 2({\bf v} \cdot {\bf b})^{2} - 1.
\end{equation}
This situation has to be compared with the one of quantum theory where only one average value can be considered for single photon measurements and it turns out to be $E_{\psi_+}({\bf a}) = E_{\psi_+}({\bf b}) = 0$, a fact which leaves no room for any statement concerning the physical properties of the individual photons. Leggett himself has proved that his proposal, as well as a more refined one considering also elliptical polarization states for the emitted photon pairs, conflicts with some predictions of quantum mechanics. Moreover, his precise model has been tested experimentally and has been rejected since the results agree with the quantum mechanical ones for cases in which Leggett's model implies a deviation from them.

Leggett's paper has stimulated a series of investigations\cite{Gisin,Groblacher,Colbeck,Parrott} aimed to prove that any crypto-nonlocal theory must produce, in the case of maximally entangled state for the two-qubits system, $f({\bf a},{\bf u}) = g({\bf b},{\bf v}) = 0$. Here, we generalize this result to any maximally entangled state in arbitrary dimensions\cite{Ghirardi2}. Even in the case of qubits, our proof is more general than those presented in the literature under several respects. First of all, we release  the request of a factorized form of the distribution function, so that we will use hidden variables $\mu$ and $\tau$ and a unique distribution function $\rho(\mu,\tau)$. Secondly, we allow the values of the local observables to depend on both sets of hidden variables, without confining their dependence, as Leggett did, to $\mu$.

\section{A general theorem on the local part of crypto-nonlocal models}

In the following, $\psi$ denotes a maximally entangled state of a system of two $N$-level systems ($N \geqslant 2$) whose Schmidt decomposition, in Dirac notation, reads
\begin{equation}\label{schmidt}
    \vert \psi \rangle = \frac{1}{\sqrt{N}} \sum_{j = 1}^N \vert v_j \rangle \otimes \vert w_j \rangle.
\end{equation}
As usual, $\{\vert v_j \rangle; j\}$ and $\{\vert w_j \rangle; j\}$ are orthonormal bases in the Hilbert spaces pertaining to Alice and Bob, respectively.
To start with, we generalize the Leggett model described in the previous section. As anticipated, we consider a general hidden variables theory, with hidden variable $\lambda = (\mu, \tau)$ distributed according to the probability density function  $\rho(\lambda) \equiv \rho(\mu,\tau)$, and the conditions of consistency with quantum mechanics, analogous to Eq. (\ref{leggcons}), read
\begin{eqnarray}\label{hvav}
    &&E_{\psi}({\bf a}) = \int A_{\psi} ({\bf a},{\bf b},\lambda) \rho (\lambda) d \lambda, \quad
    E_{\psi}({\bf b}) = \int B_{\psi} ({\bf a},{\bf b},\lambda) \rho (\lambda) d \lambda, \nonumber \\
    &&E_{\psi}({\bf a},{\bf b}) = \int A_{\psi} ({\bf a},{\bf b},\lambda) B_{\psi} ({\bf a},{\bf b},\lambda) \rho (\lambda) d \lambda,
\end{eqnarray}
where we have tacitly assumed that $\lambda$ are continuous variables. The conclusions we will draw do not depend on this assumption.

Non-locality can only be detected by comparing joint measurements performed by Alice and Bob. In fact, the average over $\lambda$ completely washes out the dependence on the setting of the other party, when a local measurement is taken into account. This fact is a necessary condition to guarantee the peaceful coexistence of the theory with the causality principle. We impose now this requirement at a different level, with the integration over $\mu$ already washing out non-locality and contextuality of local observables:
\begin{eqnarray}\label{cnlhv}
     \int A_{\psi} ({\bf a},{\bf b},\mu,\tau) \rho (\mu \vert \tau) d \mu,&=&  f_{\psi}({\bf a},\tau), \nonumber \\
     \int B_{\psi} ({\bf a},{\bf b},\mu,\tau) \rho (\mu \vert \tau) d \mu &=& g_{\psi}({\bf b},\tau),
\end{eqnarray}
where $\rho (\lambda) = \rho(\mu, \tau) = \rho (\mu \vert \tau) \rho (\tau)$. Eq. (\ref{cnlhv}) has the general form (20) for arbitrary hidden variables, and it represents the crypto-nonlocality condition. Notice that the quantum averages over $\psi$ of local observables are associated with the distribution $\rho(\tau)$, since
\begin{equation}\label{cnlhv2}
  E_{\psi}({\bf a}) = \int f_{\psi}({\bf a},\tau) \rho (\tau) d \tau, \quad
  E_{\psi}({\bf b}) = \int g_{\psi}({\bf b},\tau) \rho (\tau) d \tau.
\end{equation}
Therefore, if it were possible to access the hidden variable $\tau$, it would be possible to assign to the system, at least in principle, some meaningful local properties, in general different from the quantum ones.

A useful property is that every local map on a maximally entangled state performed on one side can be simulated by a local action performed on the other side. Making explicit reference to the orthonormal complete sets of states appearing in the Schimdt decomposition (23), we point out that for every local operator $\hat{X} = \sum_{ij} X_{ij} \, \vert v_i \rangle \langle v_j \vert$ there is an associated local operator $\hat{X}^T = \sum_{ij} X_{ji} \, \vert w_i \rangle \langle w_j \vert$ such that
\begin{equation}\label{propme}
    \hat{X} \otimes \hat{I} \, \vert \psi \rangle = \hat{I} \otimes \hat{X}^T \vert \psi \rangle.
\end{equation}
Having in mind this property, we introduce the following $\psi$-dependent bases for the spaces of Hermitian operators pertaining to Alice and Bob. For Alice, we define
\begin{eqnarray}\label{basis}
    &&\hat{F}_{ii} = \sqrt{N} \vert v_i \rangle \langle v_i \vert, \\
    &&\hat{F}^{(+)}_{ij} = \sqrt{\frac{N}{2}} \, (\vert v_i \rangle \langle v_j \vert + \vert v_j \rangle \langle v_i \vert), \quad
    \hat{F}^{(-)}_{ij} = \sqrt{\frac{N}{2}} \, (\vert v_i \rangle \langle v_j \vert - \vert v_j \rangle \langle v_i \vert), \nonumber
\end{eqnarray}
with $i < j = 1, \ldots, N$. We recast these Hermitian operators in a vector $\hat{F}$, and define the corresponding basis vector for Bob as $\hat{G} = \hat{F}^T$, by means of property (\ref{propme}). Note that, with this choice,  a generic Hermitian operator,representing a local observable can be written $\hat{A}({\bf a}) = {\bf a} \cdot \hat{F}$,
if it represents a measurement performed by Alice, or $\hat{B}({\bf b}) = {\bf b} \cdot \hat{G}$, for Bob measurements. In this context, the measurement settings are represented by $N^2$-dimensional real vectors ${\bf a}$ and ${\bf b}$ \footnote{Notice that this representation, when $N = 2$, is not equivalent to the usual expansion of a traceless Hermitian operator on the Pauli operators, which we use in the other sections. Nonetheless, ${\bf a}$ and ${\bf b}$ have a completely analogous meaning.} With these assumptions, the joint quantum averages assume the particularly simple form
\begin{equation}\label{avjo}
    E_{\psi} ({\bf a},{\bf b}) = {\bf a} \cdot {\bf b},
\end{equation}
and moreover
\begin{equation}\label{avsq}
     \langle \psi \vert \hat{A}({\bf a})^2 \vert \psi \rangle = \Vert {\bf a} \Vert^2, \quad \langle \psi \vert \hat{B}({\bf b})^2 \vert \psi \rangle = \Vert {\bf b} \Vert^2.
\end{equation}
We notice that, if ${\bf a} = {\bf b}$ and $E_{\psi}({\bf a}) = E_{\psi}({\bf b}) = 0$, the observables $A({\bf a})$ and $B({\bf b})$ are perfectly correlated. Viceversa, if ${\bf a} = -{\bf b}$ they are perfectly anti-correlated.

We consider now an arbitrary  observable $A({\bf a})$ of Alice and its associated operator $\hat{A}({\bf a})$. We can decompose it in the following form:
\begin{equation}\label{gena}
    \hat{A}({\bf a}) = \alpha_0 \hat{I} + \sum_{j = 1}^{N - 1} \alpha_j \hat{A}({\bf a}_j),
\end{equation}
where $\alpha_j$ are real coefficients for all $j = 0, \ldots N - 1$ and $\{ \hat{A}({\bf a}_j); j \}$ is a set of commuting Hermitian operators, with the following property. When $N > 2$, they have spectrum $\Omega_N = \{ -1, 0, 1 \}$, with the null eigenspace (of dimension $N - 2$) as the only degenerate one; if $N = 2$, the spectrum is $\Omega_2 = \{ -1, 1 \}$ without degeneration. The set $\{ \hat{A}({\bf a}_j);j \}$ depends on ${\bf a}$, and it is not a basis for the space of Hermitian operators, which requires $N^2$ operators. Since ${\rm Tr} \, \hat{A}({\bf a}_j) = 0$ for all $j$, it follows that
\begin{equation}\label{ava}
    E_{\psi}({\bf a}) = {\rm Tr} \, \hat{A}({\bf a}) = \alpha_0.
\end{equation}
Quantum mechanics assigns a precise value to all the observables $A({\bf a})$ and $A({\bf a}_j)$, for all $j$, since they all commute. Therefore, any hidden variables theory predictively equivalent to quantum mechanics must do the same. In particular, the values of the observables  must satisfy
\begin{equation}\label{ass}
    A_{\psi} ({\bf a},{\bf b},\lambda) = \alpha_0 + \sum_{j = 1}^{N - 1} \alpha_j A_{\psi} ({\bf a}_j,{\bf b},\lambda),
\end{equation}
where $A_{\psi}({\bf a}_j,{\bf b},\lambda)$ are associated to the observables $A({\bf a}_j)$ corresponding to $\hat{A}({\bf a}_j)$.
Notice that, if $\{A({\bf a}), B({\bf b}) \}$ is not a complete set of commuting observables and $N > 2$, these values might also depend on further parameters describing the physical context. In full generality, we should use the notation $A_{\psi}({\bf a},{\bf a}^{\prime},\ldots,{\bf b},{\bf b}^{\prime},\ldots,\lambda)$ and $B_{\psi}({\bf a},{\bf a}^{\prime},\ldots,{\bf b},{\bf b}^{\prime},\ldots,\lambda)$, where ${\bf a}, {\bf a}^{\prime}, \ldots$ and ${\bf b}, {\bf b}^{\prime}, \ldots$ specify complete sets of commuting observables for the two subsystems. For sake of simplicity, we avoid this notation,  and make explicit nonlocality and not contextuality.

Following (\ref{cnlhv}), at the intermediate level of our crypto-nonlocal theory we find
\begin{equation}\label{intf}
    f_{\psi} ({\bf a}, \tau) = E_{\psi}({\bf a}) + \sum_{j = 1}^{N - 1} \alpha_j f_{\psi} ({\bf a}_j, \tau),
\end{equation}
where we have expressed $\alpha_0$ through (\ref{ava}). The same holds true also for Bob's observables, and we can write
\begin{equation}\label{intg}
    g_{\psi} ({\bf b}, \tau) = E_{\psi}({\bf b}) + \sum_{j = 1}^{N - 1} \beta_j g_{\psi} ({\bf a}_j, \tau),
\end{equation}
where $\beta_j$ and ${\bf b}_j$ are the analogues of $\alpha_j$ and ${\bf a}_j$ respectively. Our main result follows.

\begin{theorem}\label{theo1}
 Consider the maximally entangled state of two $N$-dimensional systems $\psi$ as in (\ref{schmidt}), and the local observables $A ({\bf a})$ and $B({\bf b})$. Then necessarily $f_{\psi}({\bf a}, \tau) = E_{\psi}({\bf a})$ and $g_{\psi}({\bf b}, \tau) = E_{\psi}({\bf b})$.
\end{theorem}

{\it Proof -} In Appendix A it is shown that $f_{\psi} ({\bf a}_j, \tau) = g_{\psi} ({\bf b}_j, \tau) = 0$ in (\ref{intf}) and (\ref{intg}). We conclude that $f_{\psi}({\bf a},\tau) = E_{\psi} ({\bf a})$ and $g_{\psi}({\bf b},\tau) = E_{\psi} ({\bf b})$ for arbitrary observables.

Therefore, no crypto-nonlocal theory can assign local specifications which differ from the quantum ones. Why then further consider crypto-nonlocal theories? Is it possible to exhibit crypto-nonlocal schemes which are predictively equivalent to quantum mechanics? It turns out that these models do indeed exist, and this  opens the way for a discussion concerning  the fact that quantum mechanics does not maximally violates nonlocality. This is dealt with in the next section.

\section{Superquantum correlations in crypto-nonlocal models}

To exhibit an explicit crypto-nonlocal theory, we modify the simple nonlocal hidden variable model introduced by Bell in his famous work\cite{Bell2}. We make it more symmetric for what concerns its nonlocal features, and define the relevant variables which make it crypto-nonlocal. The model describes a pair of qubits in the singlet state $\phi_-$, defined in (\ref{singlet}), by means of the hidden variable $\lambda$, which is a unit vector in the three-dimensional real space. We assume that $\lambda$ is uniformly distributed over the unit sphere, and uniquely specified by polar angles $\mu$ and $\tau$ which we unconventionally choose to take values in the intervals $\mu \in [0,2\pi)$, $\tau \in [0,\pi)$. Such variables are related to the standard polar angles $\theta$ and $\phi$ according to \footnote{Note that, with this functional change, the surface element of the sphere turns out to be $d \Omega = \vert \sin{\mu} \vert d \mu d \tau$}:
\begin{eqnarray}
  &\mu = \theta, \; \tau = \phi \quad &{\rm for} \; y \geqslant 0; \nonumber \\
  &\mu = 2 \pi - \theta, \; \tau = \phi - \pi \quad &{\rm for} \; y < 0.
\end{eqnarray}
The assignment of $\lambda = (\mu, \tau)$ uniquely determines the nonlocally possessed values of $A({\bf a}) = {\bf a} \cdot \sigma$ and $B({\bf b}) = {\bf b} \cdot \sigma$ according to:
\begin{equation}\label{defva}
A_{\psi}({\bf a},{\bf b},\mu,\tau) = {\rm sgn} (\hat{\bf a} \cdot \lambda), \qquad
B_{\psi}({\bf a},{\bf b},\mu,\tau) = - {\rm sgn}  (\hat{\bf b} \cdot \lambda).
\end{equation}

\noindent The vectors $\hat{\bf a}$ and $\hat{\bf b}$ lie in the plane identified by ${\bf a}$ and ${\bf b}$, and are obtained from these vectors by rotating them in such a way that they are still symmetrically disposed with respect to the bisector of the angle $\omega$ (with $0 \leqslant \omega \leqslant \pi)$ between ${\bf a}$ and ${\bf b}$, and $\hat{\bf a}$ and $\hat{\bf b}$ form an angle $\hat{\omega}$ satisfying, as in the case of Bell's model, $\hat{\omega} = \pi \sin^2{\frac{\omega}{2}}$. Notice that $\hat{\omega} \leqslant \omega$ when $\omega \leqslant \pi/2$, and $\hat{\omega} > \omega$ when $\omega > \pi/2$.

By using the relations (\ref{hvav}) it is possible to check that this model is predictively equivalent to quantum mechanics (the proof matches that of the original Bell's model\cite{Bell2}). Moreover, it turns out that our model is crypto-nonlocal, with the variables $\mu$ and $\tau$ playing the role of lower and upper level hidden variables, respectively. In fact, from (\ref{defva}) it follows that every observable assumes the values $+1$ and $-1$ in opposite hemispheres of the unit sphere of $\lambda$, which has uniform distribution. Integration over $\mu$ means integration over a maximal circle, therefore $f_{\psi} ({\bf a}, \tau) = g_{\psi} ({\bf b}, \tau) = 0$, which is a necessary and sufficient condition for crypto-nonlocality, in this case.

To proceed, we evaluate the averages of the correlation functions on the lower level variable $\mu$, i.e., expressions of the type:
\begin{equation}
E_{\psi,\tau}({\bf a},{\bf b}) = \frac{1}{4} \int_{0}^{2 \pi} A_{\psi}({\bf a},{\bf b},\mu,\tau) B_{\psi} ({\bf a},{\bf b},\mu,\tau) \vert \sin{\mu} \vert d \mu.
\end{equation}

For our purposes of making explicit the superquantum character of the model at the level of the variables $\mu$, we  limit, for simplicity, our attention to four directions which, in a fixed $(x, y, z)$ reference frame, lay in the ($x,z$)-plane and are given by
\begin{eqnarray}
&&{\bf a} = (\sin{\alpha}, 0, \cos{\alpha}), \,\, {\bf a}^{\prime} = (- \sin{3 \alpha}, 0, \cos{3 \alpha}), \\
&&{\bf b} = (- \sin{\alpha}, 0, \cos{\alpha}), \,\, {\bf b}^{\prime} = (\sin{3 \alpha}, 0, \cos{3 \alpha}), \nonumber
\end{eqnarray}
with $\alpha \in [0, \pi/4]$, determining dichotomic observables $A({\bf a})$, $A({\bf a}^{\prime})$, $B({\bf b})$, $B({\bf b}^{\prime})$, as usual. This reduced scenario is sufficient to study both quantum and superquantum nonlocality. We find that
\begin{equation}\label{aver0}
    E_{\psi,\tau}({\bf a},{\bf b}) = 2 \vert \chi_1 \vert -1, \qquad E_{\psi,\tau}({\bf a}^{\prime},{\bf b}^{\prime}) = 2 \vert \chi_2 \vert -1,
\end{equation}
and
\begin{equation}\label{aver}
  E_{\psi,\tau}({\bf a},{\bf b}^{\prime}) = E_{\psi,\tau}({\bf a}^{\prime},{\bf b}) = \vert \chi_3 - \chi_4 \vert - 1
\end{equation}
when $0 \leqslant \alpha \leqslant \tilde{\alpha}$, and
\begin{equation}\label{aver2}
  E_{\psi,\tau}({\bf a},{\bf b}^{\prime}) = E_{\psi,\tau}({\bf a}^{\prime},{\bf b}) = 1 - \vert \chi_3 + \chi_4 \vert
\end{equation}
when $\tilde{\alpha} < \alpha \leqslant \pi/4$, In the above equations the functions $\chi_{j}$ are given by:
\begin{equation}\label{expr1}
    \chi_j = \chi_j (\alpha,\tau) = \frac{2 \, \cos{\tau}}{\sqrt{\cos^2{\tau} + \cot^2{\frac{\gamma_j(\alpha)}{2}}}},
\end{equation}
and the functions $\gamma_j(\alpha)$ are
\begin{eqnarray}\label{gamma}
  &&\gamma_{1} (\alpha) = \pi \sin^2{\alpha}, \quad \gamma_{2} (\alpha) = \pi \sin^2{3 \alpha}, \\
  &&\gamma_{3} (\alpha) =  4 \alpha + \pi \sin^2{\alpha}, \quad \gamma_{4} (\alpha) =  4 \alpha - \pi \sin^2{\alpha}, \nonumber
\end{eqnarray}
and $\tilde{\alpha} \simeq 0.562$ is the solution of $4 \alpha + \widehat{2 \alpha} = \pi$, where $\widehat{2 \alpha} = \pi \sin^2{\alpha}$.
We finally combine the joint correlations in the usual way,
\begin{equation}\label{f3}
    F_{\psi,\tau} \equiv E_{\psi,\tau}({\bf a},{\bf b}) + E_{\psi,\tau}({\bf a},{\bf b}^{\prime}) + E_{\psi,\tau}({\bf a}^{\prime},{\bf b}) - E_{\psi,\tau}({\bf a}^{\prime},{\bf b}^{\prime}),
\end{equation}
leading to
\begin{equation}\label{finalf}
     F_{\psi,\tau} = \left\{
                       \begin{array}{ll}
                         2 (\vert \chi_1 \vert - \vert \chi_2 \vert + \vert \chi_3 - \chi_4 \vert - 1), & \quad \alpha \in [0, \tilde{\alpha}], \\ \\
                         2 (\vert \chi_1 \vert - \vert \chi_2 \vert - \vert \chi_3 + \chi_4 \vert + 1), & \quad \alpha \in [\tilde{\alpha}, \frac{\pi}{4}].
                       \end{array}
                     \right.
\end{equation}
It is possible to check that averaging $F_{\psi,\tau}$ over the remaining variable $\tau$ consistently produces the quantum mechanical expression $F_{\psi} = - 1 - 2 \cos{\alpha} + \cos{2 \alpha}$. This is a consequence of the fact that our model is completely equivalent to quantum mechanics when averaging over $\lambda$.

\begin{figure}[t]
\centering
%\begin{center} % Requires \usepackage{graphicx}
  \includegraphics[width=8cm]{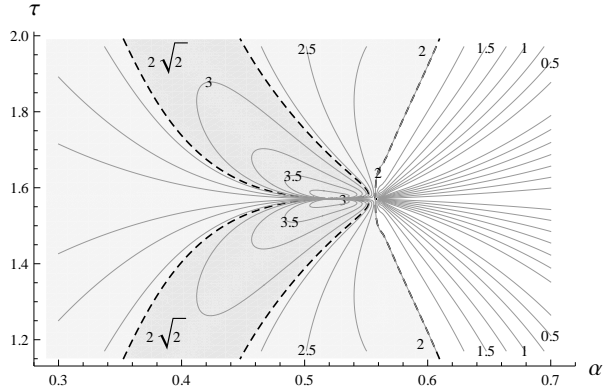} \\
 \caption{Contour plot for the function $F_{\psi, \tau}$ in the space $(\alpha, \tau)$. In the light-gray region, the model overcomes the CHSH bound $2$. In the dark-gray region, the Tsirelson bound $2 \sqrt{2}$ is violated. In this region, superquantum correlations are exhibited, and the upper bound $4$ is approached when $(\alpha, \tau) \rightarrow (\pi/6, \pi/2)$. In this limit, our model simulates, to any desired approximation, the PR box.
}\label{fig1}
%\end{center}
\end{figure}

The function $\vert F_{\psi, \tau} \vert$ is smooth, exception made for two isolated singular points at $(\alpha, \tau) = (\pi/6, \pi/2)$ and $(\tilde{\alpha}, \pi/2)$. In the neighborhoods of the first point the upper bound is approached,
\begin{equation}\label{supF}
    \sup \vert F_{\psi, \tau} \vert = 4.
\end{equation}
The simple model that we have presented is able reproduce any possible form of nonlocality which can appear in a bipartite system with two-dimensional subsystems. We can partition the space of parameters $(\alpha, \tau)$ in three regions, corresponding to locality, quantum nonlocality, and superquantum nonlocality\footnote{Note that in the region of superquantum nonlocality, for any fixed value of $\tau$ we have a model violating Bell's upper bound by an amount which depends on the chosen $\tau$ (and $\alpha$)}, see Fig. \ref{fig1}. The model is crypto-nonlocal, therefore it is consistent with relativistic causality at the intermediate level of the hidden variables. It is able to asymptotically reproduce the PR device, which is otherwise introduced as an abstract mathematical concept. As such, it provides an interesting connection between crypto-nonlocal hidden variables models and superquantum correlations. To the best of our knowledge, the only model which has been described for the simulation of the PR box by using concrete systems is based on the use of postselection\cite{Marcovitch}.

\section{Conclusions}

In this work, we have analyzed some relevant facts concerning the class of crypto-nonlocal hidden variables theories. First of all, in Theorem \ref{theo1}, we have extended a result, formerly derived for pairs of two-level systems in a maximally entangled state, using a completely different approach, to maximally entangled states of systems with arbitrary dimension. Also in the two-dimensional case our derivation differs from those in the literature.
%In particular, our result holds true also when the distribution of the hidden variables depends on the physical settings of Alice and Bob, that is $\rho(\lambda) = \rho_{{\bf a},{\bf b}}(\lambda)$ (the details are described in Appendix B). We do not describe here what this scenario could possibly represent.
Then, we have presented a simple crypto-nonlocal model for the singlet state of two qubits, which, differently from Leggett's model, is predictively equivalent to quantum mechanics. This example puts in evidence that, despite the aforementioned result, legitimate crypto-nonlocal models do indeed exist. Our model paves the way to the study of superquantum correlations at the intermediate level of crypto-nonlocal theories, and it embodies all the features which could be observed in this context. In fact, by varying the parameters which appear in the scheme, it is possible to obtain an arbitrary amount of nonlocality consistent with the requirement of relativistic causality (which follows from the conditions of crypto-nonlocality). As an asymptotic case, the extremal case of the PR box is obtained.

To the best of our knowledge, this is the first instance of a crypto-nonlocal theory which is able to reproduce the PR device. One could argue that the model\cite{Cerf} presented in Section 3 (see Eqs. (15) and (16)), in which the singlet state correlations are simulated by means of the PR box, plus additional hidden variables, has the same features of our model. In this case, one should prove that such a model is crypto-nonlocal, that it produces at the intermediate level superquantum correlations, and that it reduces to the PR device when its parameters are suitably fixed. We don't believe this is an easy task to accomplish.

The study of the specific form that nonlocality takes in the microscopic world is a fundamental step for the identification of the physical principles which are at the basis of quantum mechanics (see \cite{Pawlowski} for recent developments).  We believe that the consideration of crypto-nonlocal models might help in deepening the understanding of this relevant point.

\section*{Acknowledgements}

One of us, R.R., acknowledges the financial support by the R. Parisi Foundation. This research is partially supported by the ARO MURI grant W911NF-11-1-0268.

\appendix{Proof of Theorem 1}

We start by considering observables $A({\bf a})$ and $B({\bf b})$ with $E_{\psi} ({\bf a}) = E_{\psi} ({\bf b}) = 0$. If we choose ${\bf b} = {\bf a}$, the observables $A({\bf a})$ and $B({\bf a})$ are perfectly correlated and they have the same spectrum, then $A_{\psi} ({\bf a},{\bf a},\mu,\tau) = B_{\psi} ({\bf a},{\bf a},\mu,\tau)$. Averaging over $\mu$ leads to $f_{\psi}({\bf a}, \tau) = g_{\psi}({\bf a}, \tau)$. A similar argument holds when ${\bf b} = -{\bf a}$: $A({\bf a})$ and $B(-{\bf a})$ are perfectly anti-correlated, and it is found that $f_{\psi}({\bf a}, \tau) = -g_{\psi}(-{\bf a}, \tau)$. By combining the two relations, we conclude that $f_{\psi}(-{\bf a}, \tau) = - f_{\psi}({\bf a}, \tau)$, and similarly for $g_{\psi}$.

We further restrict our attention to operators with the spectrum $\Omega_N=\{-1,0,+1\}$, $N \geqslant 2$, and consider Alice observables for simplicity. We find convenient to write ${\mathcal H}_A = {\mathcal K} \oplus {\mathcal L}$, where ${\mathcal K}$ is the kernel of $\hat{A}(a)$, and ${\mathcal L}$ its orthogonal space, with ${\rm dim} \, {\mathcal L} = 2$. Of course, if $N = 2$, ${\mathcal H} = {\mathcal L}$. We observe that
\begin{equation}\label{pama}
    \hat{A} ({\bf a})\vert_{\mathcal L} = \tilde{\bf a} \cdot \hat{\sigma}, \quad \hat{A} (\bf a)\vert_{\mathcal K} = \hat{0},
\end{equation}
where $\hat{\sigma}$ is the vector of Pauli matrices acting on ${\mathcal L}$, and $\hat{0}$ is the null operator acting on ${\mathcal K}$. The vector $\tilde{\bf a}$ is a $3$-dimensional real vector of unit length. Notice that this vector, as well as the subspaces ${\mathcal K}$ and ${\mathcal L}$, depend on ${\bf a}$ and are univocally determined by it.

We now define a family of unitary operators acting on ${\mathcal H}$, $\{\hat{V}_{\theta} \in SU(N), \theta \in [0, \pi]\}$, by
\begin{equation}\label{unit}
    \hat{V}_{\theta} \vert_{\mathcal L} = \hat{U}_{\theta}, \quad \hat{V}_{\theta} \vert_{\mathcal K} = \hat{I,}
\end{equation}
where $\hat{U}_{\theta} \in SU(2)$ acts on ${\mathcal L}$. We assume that $\hat{U}_{\theta} = e^{i \frac{\theta}{2} \tilde{\bf c} \cdot \hat{\sigma}}$, where $\tilde{\bf c}$ is a $3$-dimensional real vector such that $\tilde{\bf c} \cdot \tilde{\bf a} = 0$. It turns out that
\begin{equation}\label{transf}
    \hat{U}_{\theta} \, \tilde{\bf a} \cdot \hat{\sigma} \, \hat{U}_{\theta}^{\dagger} = \tilde{\bf a}(\theta) \cdot \hat{\sigma},
\end{equation}
where $\tilde{\bf a}(\theta) = R (\theta) \, \tilde{\bf a}$, and $R(\theta)$ is a $SO(2)$ rotation. We can thus define a curve in the $N^2$-dimensional real space, connecting ${\bf a}$ to $-{\bf a}$, as the one-parameter family of vectors $\gamma = \{ {\bf a}(\theta); 0\leqslant \theta \leqslant \pi \}$, where ${\bf a}(\theta)$ is given by
\begin{equation}\label{gentra}
    \hat{V}_{\theta} \hat{A}({\bf a}) \hat{V}_{\theta}^{\dagger} = \hat{A}({\bf a}(\theta)).
\end{equation}
It is possible to check that ${\bf a}(0) = {\bf a}$ and ${\hat a}(\pi) = -{\bf a}$, and, as $V_{\theta}$ is a unitary transformation, $\Vert {\bf a}(\theta) \Vert = \Vert {\bf a} \Vert$ by (\ref{avsq}), and $\hat{A} ({\bf a}(\theta))$ has spectrum $\Omega_N$, and then null trace, for every $\theta \in [0, \pi]$. Finally, we notice that the decomposition ${\mathcal K} \oplus {\mathcal L}$ is independent of $\theta$, therefore the action of $\hat{V}_{\theta}$ is non-trivial only in ${\mathcal L}$ for all $\theta$. This property implies that $\gamma$ is a planar curve.

We now define a partition on this curve. For a fixed natural number $n$, we impose $\theta_j = \frac{j}{n} \, \pi$, with $j = 0, \ldots, n$. We use the shorthand notation ${\bf a}_j = {\bf a}(\theta_j)$, where ${\bf a}(\theta) \in \gamma$. Since the curve $\gamma$ is planar, we have that
\begin{equation}\label{spacing}
    {\bf a}_{j + 1} \cdot {\bf a}_j = \Vert {\bf a} \Vert^2 \cos{\frac{\pi}{n}}, \quad j = 0, \ldots, n - 1.
\end{equation}
We consider the setups in which the measurement by Alice is determined by ${\bf a}_j$ and the one of Bob  by ${\bf a}_{j + 1}$, for $j = 0, \ldots, n - 1$. Since both $\hat{A}({\bf a}_j)$ and $\hat{B}({\bf a}_{j + 1})$ have spectrum $\Omega_N$, we can write
\begin{equation}\label{proprel}
    \vert A_{\psi}({\bf a}_j,{\bf a}_{j + 1},\lambda) - B_{\psi}({\bf a}_j,{\bf a}_{j + 1},\lambda) \vert
    \leqslant \Bigl(A_{\psi}({\bf a}_j,{\bf a}_{j + 1},\lambda) - B_{\psi}({\bf a}_j,{\bf a}_{j + 1},\lambda)\Bigr)^2
\end{equation}
for all $j$. If we multiply (\ref{proprel}) by $\rho(\lambda)$ and integrate it over $\lambda$, by considering elementary properties of integrals, and making use of (\ref{avjo}), (\ref{avsq}), (\ref{spacing}), and the crypto-nonlocality conditions (\ref{cnlhv}), we obtain
\begin{equation}\label{propel2}
    \int \vert f_{\psi}({\bf a}_j, \tau) - g_{\psi}({\bf a}_{j + 1}, \tau) \vert \rho (\tau) d \tau \leqslant \frac{4 \Vert {\bf a} \Vert^2}{N} \sin^2{\frac{\pi}{2n}}
\end{equation}
for all $j$. At the l.h.s., we use the fact that $g_{\psi} ({\bf a}_{j + 1}, \tau) = f_{\psi}({\bf a}_{j + 1}, \tau)$, and then we sum these expressions for $j = 0, \ldots, n - 1$, and use the triangle inequality to obtain
\begin{equation}\label{propel3}
    \int \vert f_{\psi}({\bf a}_0, \tau) - f_{\psi}({\bf a}_{n}, \tau) \vert \rho (\tau) d \tau \leqslant \frac{4 n \Vert {\bf a} \Vert^2}{N} \sin^2{\frac{\pi}{2n}}.
\end{equation}
Now, since ${\bf a}_0 = {\bf a}$, ${\bf a}_n = -{\bf a}$, and $f$ is skew-symmetric for ${\bf a} \rightarrow -{\bf a}$, we conclude that
\begin{equation}\label{propel4}
    \int \vert f_{\psi}({\bf a}, \tau) \vert \rho (\tau) d \tau \leqslant \frac{2 n \Vert {\bf a} \Vert^2}{N} \sin^2{\frac{\pi}{2n}}.
\end{equation}
Note that, for $n \rightarrow + \infty$, the expression at the r.h.s. of (23) vanishes. Therefore $f_{\psi}({\bf a}, \tau) = 0$ almost everywhere.
This result applies as well to Bob observables.

\section*{References}

%%%%%%%%%%%%%%%%%%%%%%%%%%%%%%%%%%%%%%%%%%%%%%%%%%%%%%%%%%%%%%%%%%%%%%%%


\begin{thebibliography}{0}

\bibitem{Erwin} E. Scr\"{o}dinger, Proc. Camb. Philos. Soc. 31, 555 (1935)

\bibitem{EPR}  A. Einstein, N. Rosen and B. Podolsky, Phys. Rev. 47, 777 (1935)

\bibitem{von} J. von Neumann, {\it Mathematische Grundlagen der Quantenmechanik}, Springer, Berlin, 1932

\bibitem{Bell1} J.S. Bell, Rev. Mod. Phys., 38, 447 (1966)

\bibitem{Bohm} D. Bohm, Phys. Rev. 85, 166 (1952), {\it ibidem} 180 (1952)

\bibitem{Bell2} J.S. Bell, Physics 1, 195 (1964)

\bibitem{CHSH} J.F. Clauser, M.A. Horne, A. Shimony and R.A. Holt, Phys. Rev. Lett. 23,  880 (1969)

\bibitem{Popescu} S. Popescu and D. Rohrlich, Found. Phys. 24, 379  (1994)

\bibitem{Jarrett} J. Jarrett, {\it Nous}, {\bf 18}, 569 (1984).

\bibitem{Shimony} A. Shimony, in: {\it Foundations of Quantum Mecvhanics in the Light of New Technology}, p. 225, S. Kamefuchi et al. (eds.), Physical Society of Japan, Tokyo (1984)

\bibitem{Leggett} A.J. Leggett, Found. Phys. 33, 1469 (2003)

\bibitem{Ghirardi1} G.C. Ghirardi and R. Romano, arXiv:1203.3093

\bibitem{Kochen} S. Kochen and E.P. Specker, Jour. of Math. and Mech. 17, 59 (1967)










%\bibitem{Suppes} P. Suppes and  M. Zanotti, in: {\it Logic and Probability in Quantum Mechanics}, p. 445,  P. Suppes (ed.), Reidel, Dordrecht (1976).




\bibitem{Tsirelson} B.S. Cirel'son, Lett. Math. Phys. 4, 93 (1980)


\bibitem{Cerf} N.J. Cerf, N. Gisin, S. Massar and S. Popescu, Phys. Rev. Lett. 94, 220403 (2005)

\bibitem{Toner} B.F. Toner and D. Bacon, Phys. Rev. Lett. 91, 1879804 (2003)



%
%\bibitem{Pawlowski} M. Pawlowski, T. Paterek, D. Kaszlikowski, V. Scarani, A. Winter and M. Zukowski, Nature 461, 1101 (2009)
%
%\bibitem{Einstein} A. Einstein in: ``The Correspondence Between Albert Einstein and Max and Hedwig Born," Walker Publishing Group, 1971
%
%\bibitem{Popescu} S. Popescu and D. Rohrlich, Found. Phys. 24, No. 3, 379 (1994)
%
%\bibitem{Barrett} J. Barrett and S. Pironio, Phys. Rev. Lett. 95, 140401 (2005)
%
%\bibitem{Cerf} N.J. Cerf, N. Gisin, S. Massar and S. Popescu, Phys. Rev. Lett. 94, 220403 (2005)
%
%\bibitem{Skrzypczyk} P. Skrzypczyk, N. Brunner and S. Popescu, Phys. Rev. Lett. 102, 110402 (2009)

\bibitem{Gisin} C. Branciard, N. Brunner, N. Gisin, C. Kurtsiefer, A. Lamas-Linares, A. Ling and V. Scarani, Nature Physics 4, 681 (2008)

\bibitem{Groblacher} S. Groblacher, T. Paterek, R. Kaltenbaek, C. Bruckner, M. Zukowski, M. Aspelmeyer and A. Zeilinger, Nature 446, 871 (2007)

\bibitem{Colbeck} R. Colbeck, and R. Renner, Phys. Rev. Lett. 101 050403 (2008)

\bibitem{Parrott} S. Parrott, arXiv:0707.3296

\bibitem{Ghirardi2} G.C. Ghirardi and R. Romano, arXiv:1203.4687

\bibitem{Marcovitch} S. Marcovitch, B. Reznik and L. Vaidman, Phys. Rev. A 75, 022102 (2007)

\bibitem{Pawlowski} M. Pawlowski, T. Paterek, D. Kaszlikowski, V. Scarani, A. Winter and M. Zukowski, Nature 461, 1101 (2009)


%\bibitem{Marcovitch} S. Marcovitch, B. Reznik and L. Vaidman, Phys. Rev. A 75, 022102 (2007)
%
%\bibitem{Tsirelson} B.S. Cirel'son, Lett. Math. Phys. 4, 93 (1980)
%
%


%\bibitem{Colbeck2} R. Colbeck and R. Renner, Nature Communications 2, 411 (2011)
%
%\bibitem{Ghirardi} G.C. Ghirardi and R. Romano, {\it Quantum and Superquantum correlations}, submitted for publication


\end{thebibliography}
\end{document}